\documentclass[12pt,preprint]{aastex}

\begin{document}

\title{Spatial Variations of Galaxy Number Counts in the Sloan Digital Sky Survey. II. Test of Galactic Extinction in High Extinction Regions}

\author{Naoki Yasuda, Masataka Fukugita}

\affil{Institute for Cosmic Ray Research, University of Tokyo, Kashiwa, 277-8582, Japan}
\email{yasuda@icrr.u-tokyo.ac.jp}

\and
\author{Donald P. Schneider}
\affil{Department of Astronomy and Astrophysics, 525 Davey Laboratory, Pennsylvania State University, University Park, PA 16802}

\begin{abstract}

Galactic extinction is tested using galaxy number counts at low
Galactic latitude obtained from five band photometry of the
Sloan Digital Sky Survey. The spatial variation of galaxy
number counts for low extinction regions of $E(B-V) < 0.15$ is
consistent with the all-sky reddening map of \citet*{SFD} and the
standard extinction law. For higher extinction regions of $E(B-V) >
0.15$, however, the map of \citet{SFD} overestimates the reddening
by a factor up to 1.4, which is likely ascribed to the departure from
proportionality of reddening to infrared emissivity of dust. This result is
consistent with the analysis of \citet{AGa} for the Taurus dark cloud
complex.

\end{abstract}


\defcitealias{PaperI}{Paper~I}
\defcitealias{SFD}{SFD}

\section{Introduction}

The Sloan Digital Sky Survey \citep[SDSS:][]{York2000}
provides an unprecedented wide-field imaging survey of the sky 
with homogeneous, few-percent-error photometry. It allows
study of the spatial variation of the projected galaxy number density over
the wide field. In our previous publication 
\citep[Paper~I]{PaperI}, it was shown
that the variation of the galaxy number counts is accounted 
for by the Galactic extinction and large-scale galaxy clustering.

Galaxy counts provide a clean test of the Galactic extinction
integrated out to the edge of the Milky Way, provided that the
galaxies are sampled to sufficiently faint magnitude and their numbers
are sufficiently large \citep[see, e.g.,][]{Hubble,BH}. The test does not
need any models or external calibrations to derive extinction. Moreover, using
SDSS photometric data, extinctions in five-bands can be determined
independently and the extinction curve can be derived. The primary limitation
of this technique is that one cannot explore the small scales where spatial variations
of the galaxy distribution are strongly affected by galaxy clustering. The mean amount of 
Galactic extinction can be well documented, and the error for scale
larger than a degree can be made smaller than the order that concerns
us. The study in \citetalias{PaperI} indeed verified the validity of
the reddening map of \citet[SFD]{SFD} on the scale of a degree and the
extinction laws of \citet{Cardelli} and \citet{ODonnell} to within 5\%
for five optical colours. More recently, a similar but more detailed
analysis reveals in very
low extinction regions slight underestimates of extinction in the SFD map,
which are likely due to the contamination of infrared emission
from galaxies in the map of dust far infrared emission \citep{Yahata}. 

Our study of extinction in \citetalias{PaperI} was limited to
regions of low extinction [$E(B-V) < 0.15$] since the galaxies
used are derived from the main survey area of SDSS \citep[][DR1]{DR1}
that intentionally avoids the region of large extinction to
explore extragalactic science. A significant amount of SDSS imaging data,
however, has been obtained at low Galactic latitudes for
the purpose of photometric commissioning and calibrations. A part
of such surveys, that include the Orion region, are published in
\citet{OrionDR}. The purpose of the present paper is to extend our
study to high extinction regions and examine the validity of the
reddening map and the extinction law given by \citetalias{SFD}.

We note that some indication was already reported that the SFD
reddening map may not be correct for high extinction regions from a
study of stars in the background of the Taurus dark cloud complex
\citep{AGa}. Arce \& Goodman suggested that extinction derived from SFD is
overestimated by a factor of $1.3-1.5$ in regions 
with $A_V > 0.5$ mag, or $E(B-V) > 0.16$. Our study will test
this result in the Orion region.

In Section 2 we describe the sample, data selection and the procedure
to derive extinction from galaxy number counts.  Section 3 presents
the result together with some discussion.  A summary is given in
Section 4.

\section{Data and analysis}
\label{sec:data}

\begin{figure}
\plotone{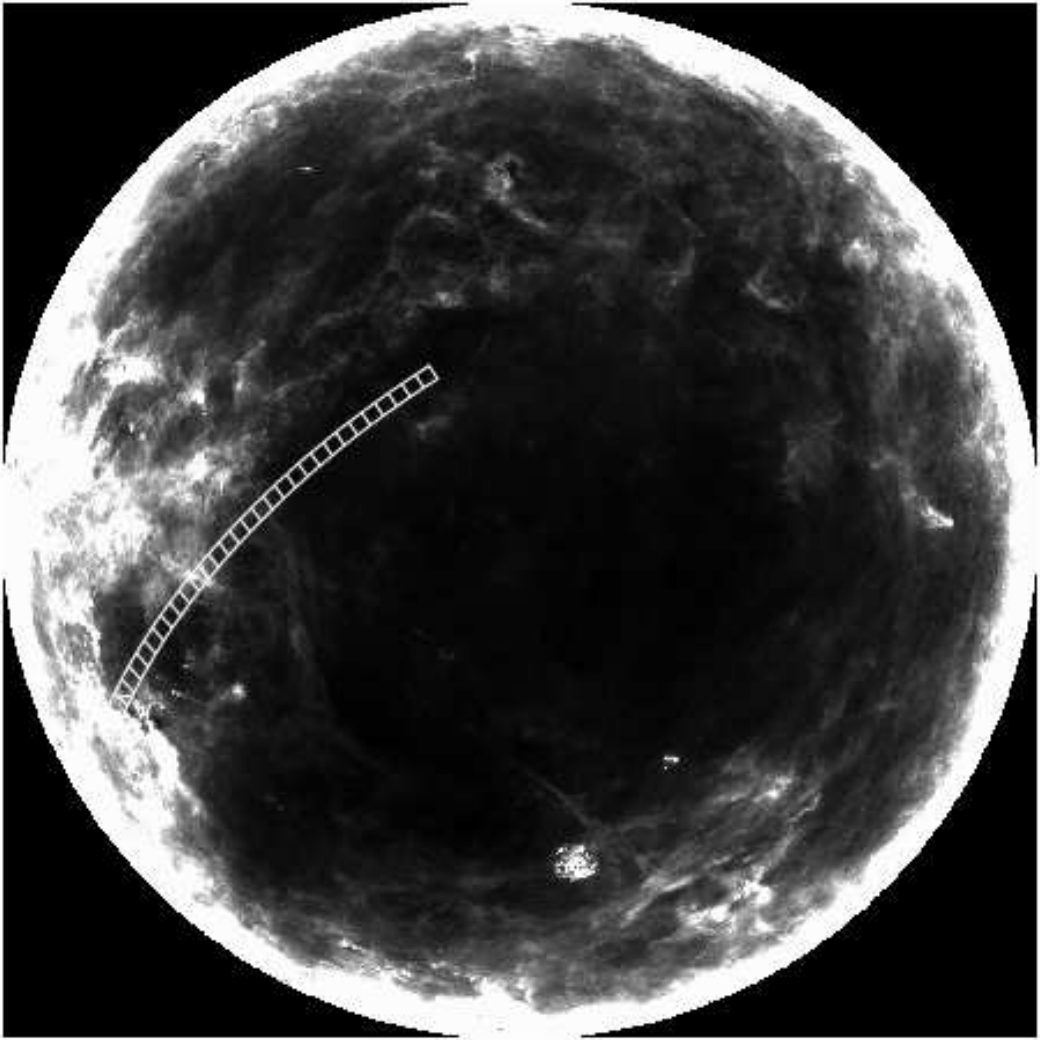}
\caption{Region for the current analysis overlaid on the reddening map
of SFD in the Galactic southern hemisphere with the pole at the centre
($b=-90\arcdeg$). The left edge corresponds to $l=180\arcdeg$ and the
top is $l=90\arcdeg$. The stripe runs from $(l,b) = (125\arcdeg, -
62\arcdeg)$ to $(207\arcdeg, -10\arcdeg)$ and the square along the
stripe represents $2.5\arcdeg$ square regions. 
The progression of black to white in the figure corresponds to
increasing values of $E(B-V)$.
\label{fig:map}}
\end{figure}

\begin{figure}
\plotone{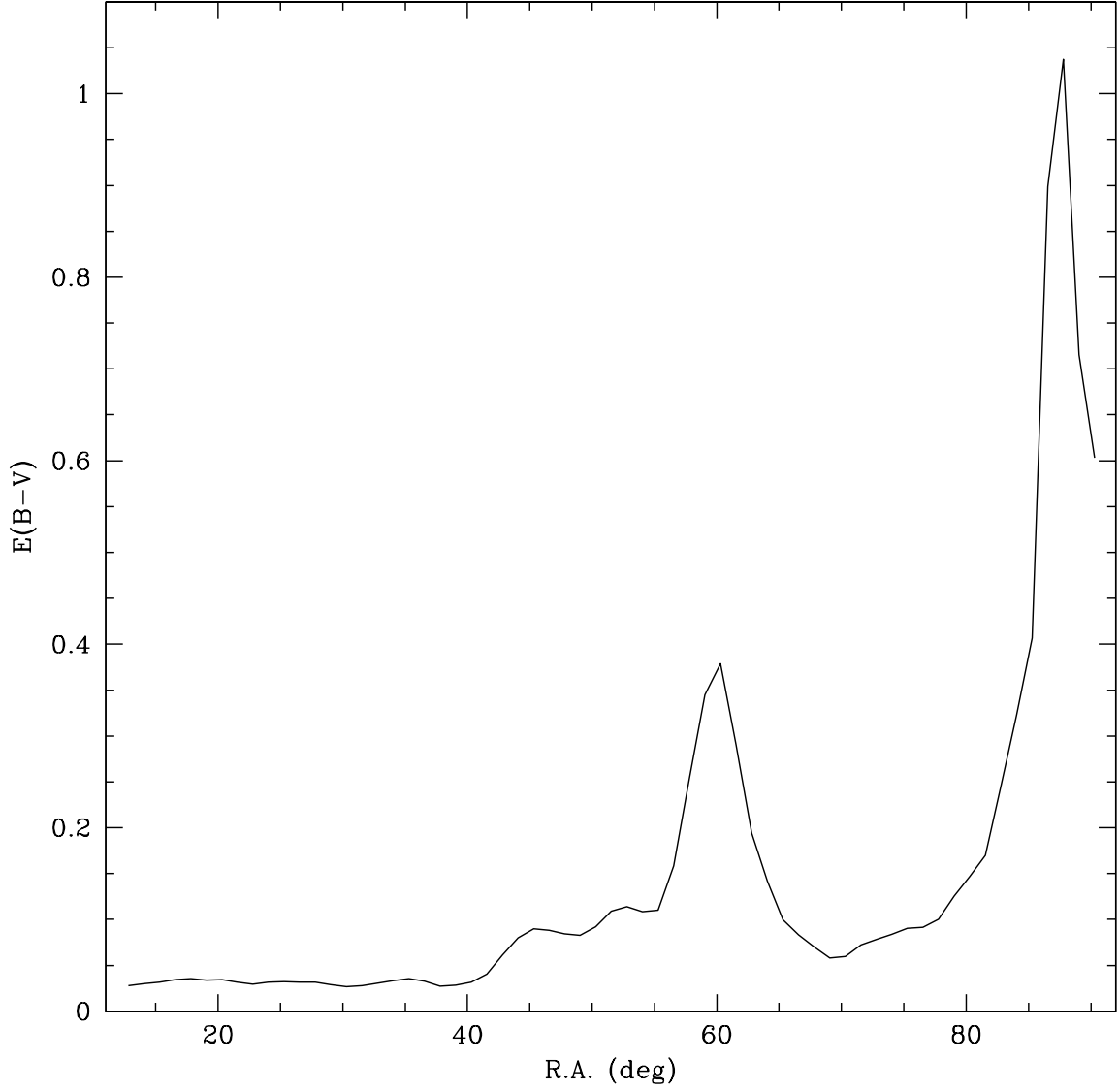}
\caption{Variation of the mean reddening $E(B-V)$ of SFD averaged over
the $2.5\arcdeg$ square region along the stripe in this study.
The Orion Complex is located at R.A. = $90\arcdeg$.
\label{fig:raE}}
\end{figure}

The imaging by SDSS telescope \citep{Gunn98,Gunn06} is carried out with
the SDSS $ugriz$ filters \citep{Fukugita1996}. 
Our analysis is based on the data given by \citet{OrionDR}. We 
use the data of stripe 82 (southern equatorial stripe) which are
derived from photometric scans, runs 211(82S), 259(82N), and 273(82S).
This is the only stripe in \citet{OrionDR} for which a north-south
pair of stripes, which are needed to form a contiguous
area, are observed. 
The region covers $11\fdg55 < \alpha < 91\fdg55$ and $-1\fdg25 <
\delta < +1\fdg25$, corresponding to $125\arcdeg \lesssim l \lesssim
207\arcdeg$ and $-62\arcdeg \lesssim b \lesssim -
10\arcdeg$. Figure \ref{fig:map} shows the region covered by current
data set overlaid in the reddening map of \citetalias{SFD} for the
Galactic southern hemisphere. The total area of our data set, 200
deg$^2$, is about 10\% of that covered by \citetalias{PaperI}
derived from SDSS DR1. The mean selective reddening $E(B-V)$ increases
from 0.03 at a high Galactic latitude to $\approx 1$ in the Orion
region close to the Galactic plane, as seen in Figure \ref{fig:raE},
which shows $E(B-V)$ of \citetalias{SFD} averaged
over the $2.5\arcdeg$ square regions along the stripe. We note that
extinction in dark cloud regions is patchy, and $E(B-V)$ occasionally
reaches $>10$ mag for small regions, typically for $15\arcmin$.

An object catalogue and five colour band images are given in
\citet{OrionDR}. The catalogue is created by photometric pipeline {\tt
photo} v5\_4\_25, which is used to produce Data Release 2 \citep{DR2}
and later data releases \citep{DR3,DR4,DR5}. We select galaxies primarily
using {\tt objType} parameter in the catalogue, but drop objects which
are flagged as {\tt EDGE}, {\tt BRIGHT}, {\tt SATURATED}, or {\tt
BLENDED}. It sometimes happens that some objects are not detected in
all five bands. We judge the detection using logical {\tt OR} of {\tt
BINNED1}, {\tt BINNED2}, and {\tt BINNED4} flag set per band, which
mean that objects are detected in a $1\times1$, $2\times2$, and
$4\times4$ binned image, respectively. The entries that are not
detected in a given band are not included in our number counts in that
band. There are some entries in the catalogue whose corresponding
objects appear to be much too faint for the quoted magnitudes; their
surface brightness is indeed faint. The occurrence of this type of spurious
detection is common in all colour bands. When the surface brightness
distribution is plotted for the galaxy sample, these objects produce a
spike at nearly zero surface brightness. Since the reason for the
inclusion of such entries in the catalogue is not clear, we apply a
surface brightness cut at the level of sky background to exclude
those entries in the $r$ band. This procedure automatically removes
the surface brightness spikes in the $g$, $i$ and $z$ bands. For the $u$ band, there are many
true galaxies that have low surface brightness, so we cannot separate
the fake spike from the tail of true galaxies; we simply assume that
the removal of the spike in the $r$ band also removes spurious objects in the $u$
band. This procedure rejects about
2\% of the $r$ band selected galaxies in the dereddened magnitude range of $17.5 < r <
19.5$. We find that most of these entries are flagged as {\tt
nopetro}, but if we would use {\tt nopetro} as the selection criterion
we would miss an additional 3\% of objects, which are mostly true 
galaxies.

\begin{figure}
\plotone{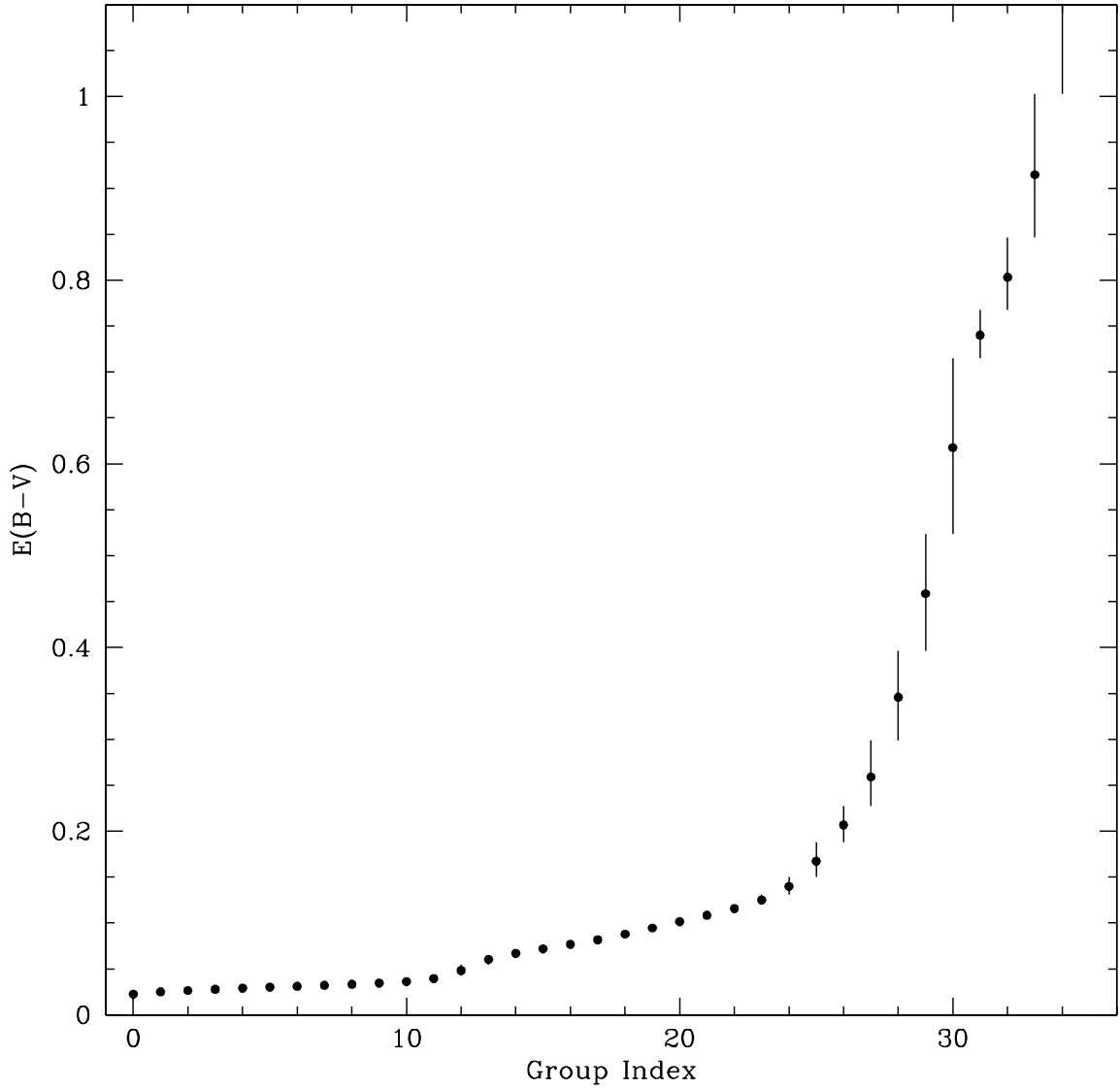}
\caption{Range of $E(B-V)$ of the SFD map for 32 groups, each group
consisting of 4225 pixels that form the effective $2.5\arcdeg$ square
region. (The four rightmost points are the groups having one fourth
the number of pixels of the others, corresponding to effective
$1.25\arcdeg$ square.) Filled circles denote the mean values and bars
show the minimum and maximum.
\label{fig:Ebv}}
\end{figure}

To account for patchy extinction structure in a heavily
reddened region, we divided the total area of 200 square degrees into
pixels of $2.3\arcmin\times2.3\arcmin$. This pixel size was adopted to
match that of the SFD reddening map ($2.372\arcmin$). There are 135,200
such pixels in our area. All pixels are sorted in the order of increasing
$E(B-V)$ values of the SFD map and grouped every 4225 pixels, so that
the sum of the area of each group amounts to $(2.5\arcdeg)^2$, which is
large enough to determine extinction from galaxy number counts.
Figure \ref{fig:Ebv} shows the mean $E(B-V)$ value of each group. Half
of the groups have $E(B-V) < 0.1$. Since the group of the largest $E(B-V)$
spans too wide a range of reddening, we subdivide it into four: namely
the last four points have been grouped with 1056 pixels (one quarter
of 4225). The last dataset includes pixels with $E(B-V)$ from 1 to
33; it is dropped from our analysis.

The present analysis differs from that in \citetalias{PaperI} in that
contiguous regions of 2.5 degree squares are considered. Such an
analysis is not appropriate to deal with the sample that includes high
extinction regions, since extinction is very patchy for dark cloud
regions and the average of extinction over a large area does not
necessarily match the magnitude offset calculated from galaxy
counts\footnote{For example, in an area where half of the
region is clear and the other half is extincted by 20 magnitudes,
then the mean extinction would be 10mag. However, galaxies can be
observed in the clear area and the number of galaxy will be half that
expected for the case without extinction. This leads us to infer that
the magnitude offset would be about 0.5mag.}. For this reason we
divide the area into small regions and assemble the region according
to extinction values in the present paper. For low extinction
regions, the details of the binning are not important. 

We use the Petrosian magnitude \citep[see][]{EDR,TS} in our study.
Petrosian magnitudes are suitable for the analysis given
in this paper, since the Petrosian radii are unaffected by the
foreground extinction and Petrosian magnitudes measure the same
fraction of the flux of galaxies regardless of their foreground
extinction. This means that we can correct the extinction properly
just by subtracting the extinction value in each band. This does not
hold for other galaxy magnitudes such as isophotal or aperture
magnitudes. 
We have to note that Petrosian magnitudes are not
corrected for seeing variations unlike model magnitudes
which are PSF-convolved model fits.
The fraction of total magnitude measured by
Petrosian magnitude will change as galaxies become smaller
as a result of worse seeing \citep{Blanton}.
This effect is only a few percent for the galaxies used in
this study ($17.5 < r < 22.0$ or $1.0'' < r_{50} < 1.3''$) and
does not affect our results.
We refer astrometric calibrations to \citet{Pier} and
photometric calibrations of the main survey to
\citet{Smith} \citep[see also][]{Hogg,Tucker} and
\citet{Ivezic}.  Since the secondary standard star ``patches'' are
sparse or nonexistent for much of the Orion region, photometric
calibrations for the current data use the \"ubercalibration
algorithm \citep[See][for details]{OrionDR,Gunn06}.

We derive the mean extinction-free galaxy number count (differential
count) $\bar{N}(m)$ from the entire sample in low extinction regions
of $E(B-V) < 0.1$ by employing the SFD reddening map and the default
standard extinction law, $k(r) = A_r/E(B-V)=2.751$
\citepalias[Table 6 of][]{SFD}. We take this
mean relation as the reference. We then count the number of galaxies for
each group without applying the extinction correction, and fit to the
reference count by shifting the amount of magnitude $\Delta m$, i.e.,
as $\bar{N}(m+\Delta m)$; this $\Delta m$ represents the extinction in
the specific region.

\begin{figure}
\plotone{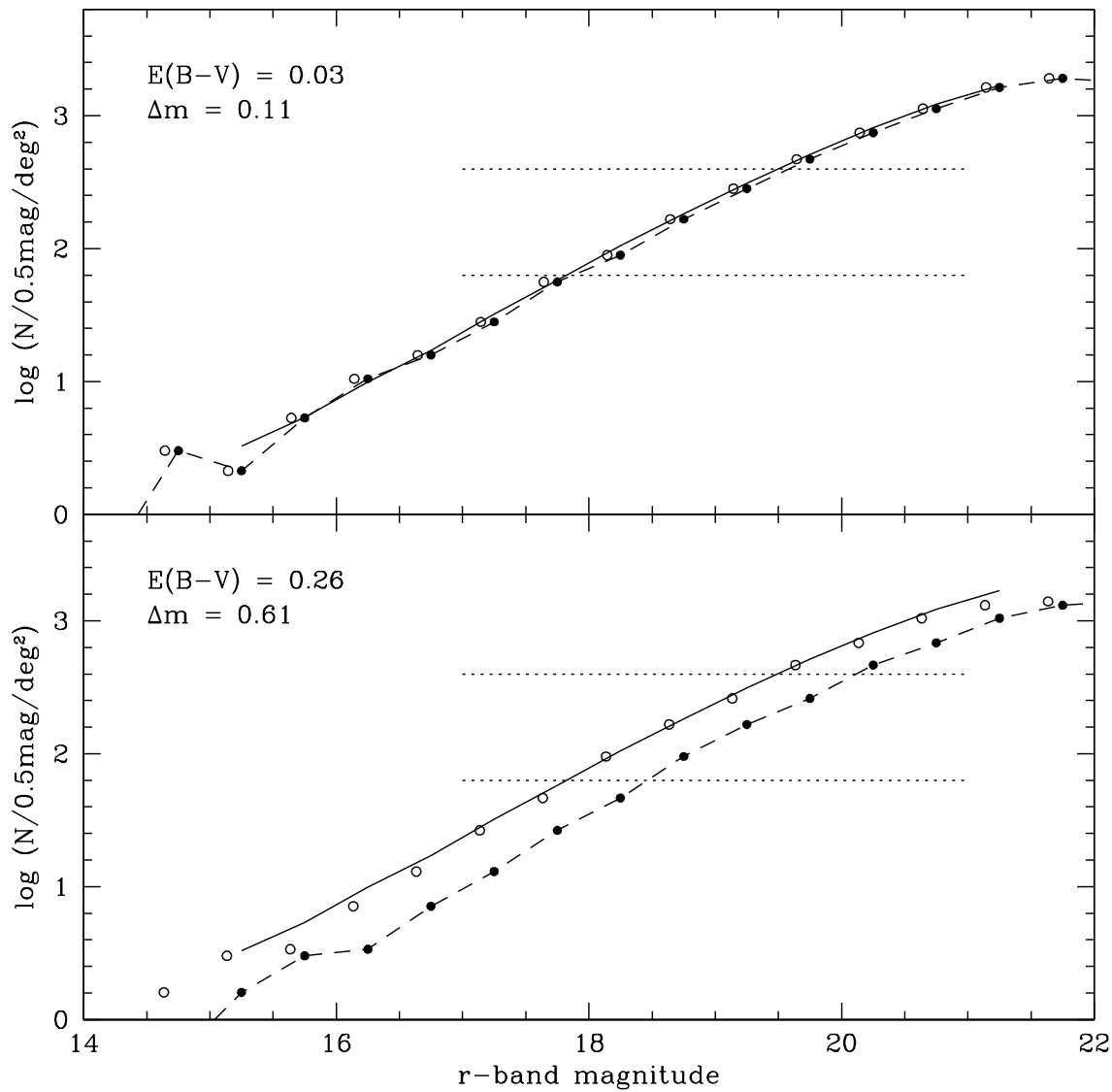}
\caption{Example galaxy number counts in the $r$-band. Upper panel is
for a low extinction ($E(B-V)=0.03$) region and lower panel is for a high
extinction ($E(B-V)=0.26$) region. Solid lines are the reference
galaxy counts. Solid points (and dashed lines) represent observed
galaxy counts, which are shifted to match the reference counts by
the amount of $\Delta m$ (denoted by open circles). The bands indicated by
dotted lines are the range that is used for matching.
\label{fig:ncfig}}
\end{figure}

We first work with the $r$ band counts, but extend the study later to
other colour bands. It is desirable to work with number counts at a
level as faint as possible, so that the galaxy number density is
sufficiently large to minimise the Poisson noise and the spatial
distribution of galaxies is sufficiently smooth to minimise the
large-scale clustering effects. \citetalias{PaperI} uses the data in
the magnitude range of $r = 18.5 - 20.5$. In the current study, this approach is not
appropriate; the reddening expected in the SFD map varies from $E(B-V)
= 0.03$ to $0.9$ (regions of the largest extinction are
discarded). This range corresponds to $A_r = 0.08 - 2.5$, and the
count in highly reddened regions falls out of the range set as the
reference magnitude band. To avoid this problem we set the range in a
way that dereddened magnitude range is the same for all regions. We use the
counts whose number density per square degrees per 0.5 mag is
\begin{equation}
1.8 < \log N(m) < 2.6,
\label{eq:Nm}
\end{equation}
and fit $N(m)$ to the reference count to derive $\Delta m$. This range
corresponds to $r = 17.5 - 19.5$ on the reference galaxy number count,
which is one mag brighter than that in \citetalias{PaperI}. This is
still reasonably faint, yet photometric measurements are made at a
high signal-to-noise ratio and star-galaxy separation is sufficiently
reliable even with extinction. This ability to classify objects is particularly
important for our study because the
contamination of stars becomes more serious at low Galactic latitude
where the star density is high and high extinction pushes the objects
to fainter magnitudes. The number of galaxies contained in
$2.5\arcdeg\times2.5\arcdeg$ area integrated over $1.8 < \log N(m) <
2.6$ is approximately 4100. The expected Poisson noise of 1.6\% is
negligible for the present work. It occasionally happens that the
faint end of eq. (\ref{eq:Nm}) goes beyond the magnitude at which
incompleteness starts ($r=22.0$); in this case we drop the data at the bins that
go beyond the incompleteness limit. Figure \ref{fig:ncfig}
shows examples of $r$-band galaxy counts in low and high extinction
regions. We see in this figure how magnitude offsets are evaluated.

The normalisation of the reference number count 
in the current sample is lower than that derived in 
\citetalias{PaperI} by 5\% ($\Delta \log N(m)=0.02$)\footnote{
It is lower by 7\% ($\Delta \log N(m)=0.03$) than \citet{Yasuda}
which is based on the SDSS Early Data Release \citep{EDR}.}.
A half of this difference is explained by the zero point offset
of photometry between DR1 and \citet{OrionDR}, the latter being fainter
by $\approx 0.02$ mag. The other half is due to the present
omission of objects that produce this spike at zero surface brightness
explained above: we did not apply this cut in \citetalias{PaperI}.

We carry out a similar analysis for four other colour bands. The
reference range of counts (as log $N$), the corresponding dereddened
magnitude range, approximate number of galaxies, and the limiting
magnitude are given in Table \ref{tbl:param}. Note that we must
choose a range of number of galaxies that is smaller in the bluer bands. In
particular, the $u$ counts go quickly out of the reference range in
the presence of extinction. After deriving
reference number counts in the extinction-free limit of the low
extinction galaxy samples, we compute $\Delta m_\lambda$ for the
counts in each group. In deriving the reference counts
we assume the standard extinction curve, $k(\lambda)=A_\lambda/E(B-V)$
with $k(\lambda)=5.155$, 3.793, 2.086, and 1.479 for $u$, $g$, $i$, and $z$.
Note that $k(\lambda)$ varies by a factor of 3.5 across $u$ to $z$,
and by a factor 2.5 if we drop the $u$ band. This makes the use of
$g,r,i,z$ colours appropriate to study the extinction curve, even if we
would exclude $u$ for its poorer photometry.
 
We find the offset in the normalisations of the reference count,
$\Delta \log N(m)=0.3,~ 0.03,~ 0.03$, and $0.05$ for $u$, $g$, $i$,
and $z$ compared to those in Paper I. The offsets in $g$, $i$, and $z$
bands are explained in the same way as for the $r$ band. The offset in
$u$ band is large. The shape of the counts at the faint end also
differs from that in paper I. These differences originate from the
fact that $u$ band surface brightness is faint for many galaxies and
the photon collecting efficiency in $u$ band is low, both contributing
to poor photometry in this colour band. A comparison of the photometry
between DR1 and \citet{OrionDR} for common galaxies shows that the two
magnitudes differ typically as much as 0.5 mag randomly\footnote{For
stars this large scatter is not seen. The difference of the two
magnitudes is no more than twice those in $r$ or $g$ bands; the mean
scatter is 0.03 mag at $u=19$.}. There are also some differences in
the selection procedure. In the present analysis we required
`detection' in the $u$ band in 1$\times$1, 2$\times$2, or 4$\times$4
binned image, which we did not do in Paper I. The surface brightness cut
in $r$ is exercised in the present analysis. The effect is small for
the $r$ band, but this drops 10\% of objects for $18.0<u<20.0$ in the
$u$ band count.  These effects altogether might induce a large error
in for the $u$ band count. We expect, however, a substantial part of
the errors are likely to cancel when we deal with relative quantities,
like those in our analysis.

\section{Results}

\begin{figure}
\plotone{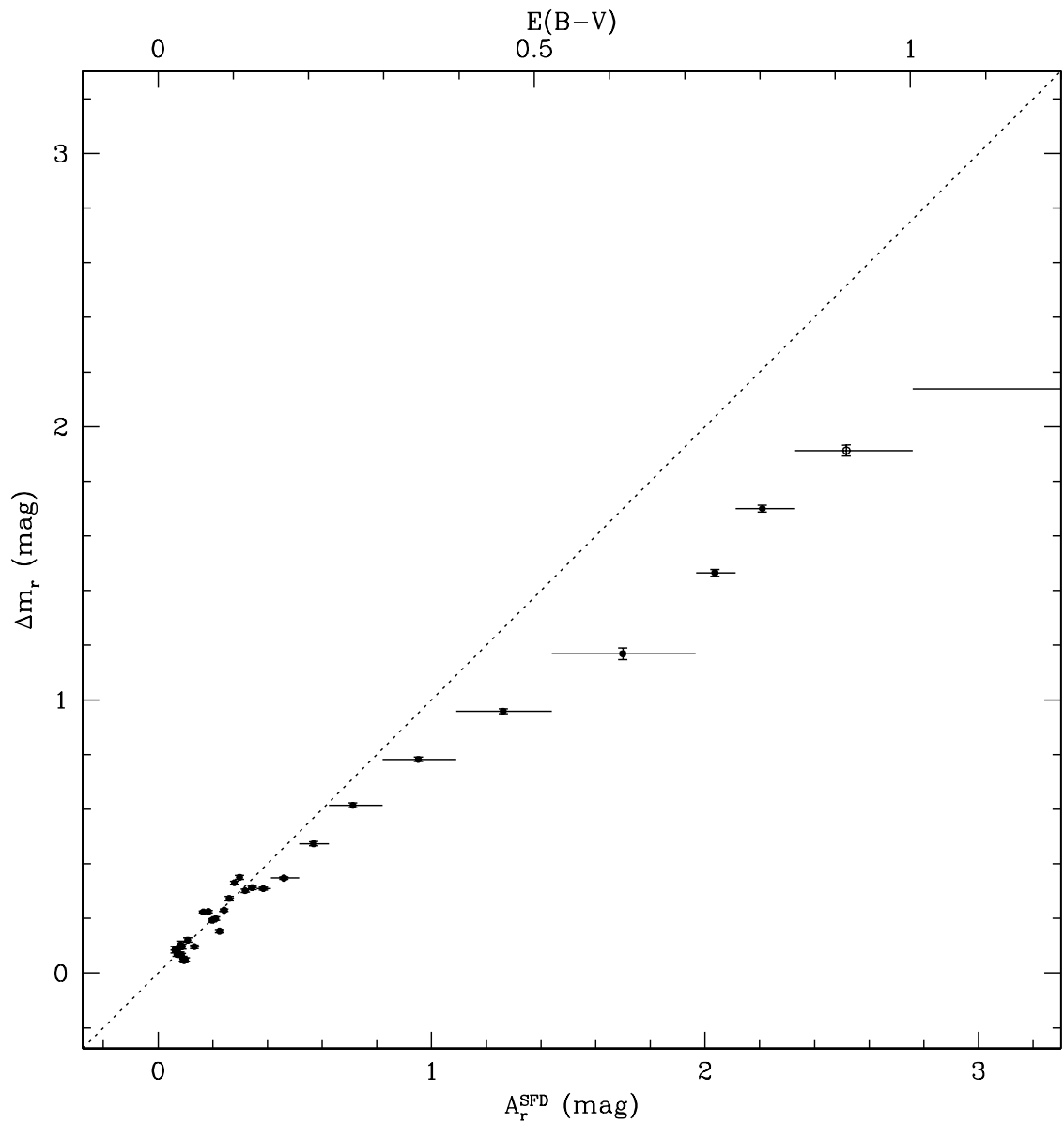}
\caption{Magnitude offsets $\Delta m_r$ corresponding to the variation
of galaxy number counts in groups corresponding to
the effective area of $(2.5\arcdeg)^2$
plotted against
mean extinction $A_r^{\rm SFD}$ calculated from the SFD reddening map
and the standard extinction curve. 
Horizontal bars show the region of $A_r$ in each group and vertical
bars are errors obtained by a jackknife estimate.
The open circles represent the
region where $A_r^{\rm SFD}$ is so large that the faint end of the
reference magnitude range falls beyond the incompleteness limit. The
dotted line is the identity regressions line $\Delta m_r
= A_r^{\rm SFD}$.  
\label{fig:Ar}}
\end{figure}

In Figure \ref{fig:Ar} we show $\Delta m_r$ versus Galactic
extinction, $A_r^{\rm SFD}$, calculated from the reddening map of
\citetalias{SFD} averaged over each group with the effective area of
$(2.5\arcdeg)^2$ assuming the standard extinction curve.  For each
group, we applied a jackknife method with the data divided into 10
samples to estimate statistical errors.  The horizontal bars stand for
the range of $A_r$ in each group.  The data points for which the faint
end of the observed magnitude corresponding to eq. (\ref{eq:Nm})
becomes fainter than the incompleteness limit are shown by open
circles. We note that the scatter of the points is substantially reduced
compared to those in \citetalias{PaperI}, where it is dominated by large
scale clustering of galaxies.
We expect a $\pm 0.11$ mag variation from large scale galaxy
clustering if the area considered were taken from contiguous regions
along the stripe. The observed scatter (0.04 mag) is consistent with
the mean error of fitting of the reference count to the data of
$N(m)$, which is 0.04.  This reduced scatter obviously arises from
the fact that the information of the large scale structure is mostly
lost by division and reassembling of the area.  This is an advantage
of the present procedure from the viewpoint of testing the effect of
extinction.

Figure \ref{fig:Ar} shows that for $A_r^{\rm SFD} < 0.4$ (or $E(B-V) <
0.15$), $\Delta m_r$ is proportional to $A_r^{\rm SFD}$ when averaged
over the scatter. This agrees with what we demonstrated in
\citetalias{PaperI}: the galaxy counts are consistent with the reddening map
and the standard extinction law. For a high extinction regime,
$A_r^{\rm SFD} > 0.4$ or $E(B-V) > 0.15$, however, the data
significantly deviate from the identity regression line: $\Delta m_r$
is smaller than $A_r^{\rm SFD}$, indicating that $A_r^{\rm SFD}$
overestimates the Galactic extinction.

This overestimate of Galactic extinction in the SFD map is consistent
with the analysis of \citet{AGa,AGb}, who studied extincion in the
Taurus dark cloud complex using the number count and reddening of
background stars and 100$\mu m$ infrared emission (calibrated with
star number counts). They reached the conclusion that the SFD map overestimates
the extinction by a factor of $1.3-1.5$ for the region of $A_V > 0.5$.

\begin{figure}
\plotone{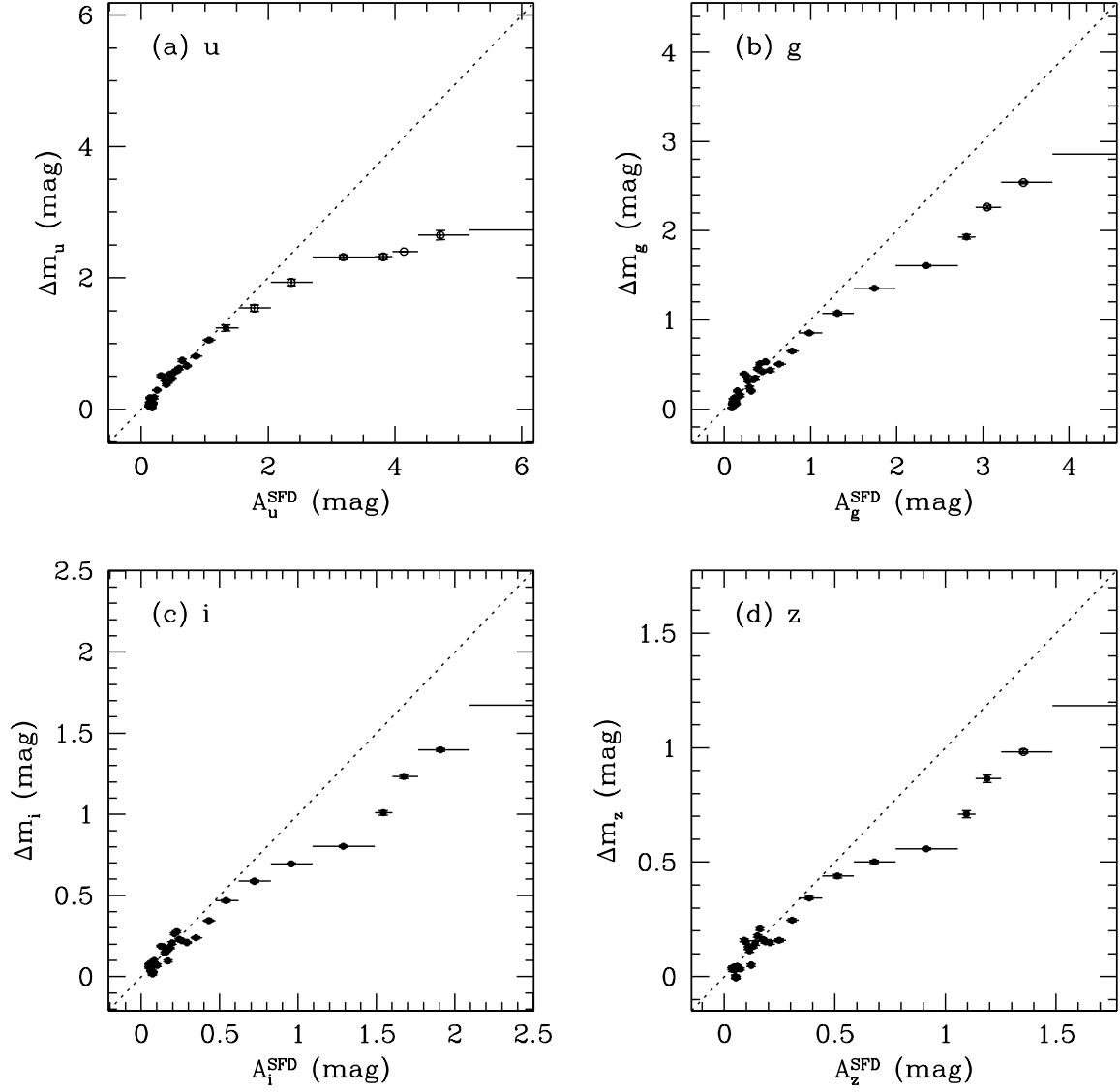}
\caption{Same as Figure \ref{fig:Ar}, but for the other four SDSS colour bands:
(a) $u$ band, (b) $g$ band, (c) $i$ band, and (d) $z$ band.
\label{fig:Augiz}}
\end{figure}

We perform a similar analysis for the $u$, $g$, $i$, and $z$ colour
bands. The relations between $\Delta m_\lambda$ and $A_\lambda^{\rm
SFD}$ are shown in Figure \ref{fig:Augiz}. For the $g$, $i$ and $z$
bands we see a similar departure from the $\Delta m_\lambda=
A_\lambda^{\rm SFD}$ line for $E(B-V)\gtrsim 0.15$, while the data for
$E(B-V)\lesssim 0.15$ are consistent with the identity line. The
departure for the $u$ band appears somewhat different, but we note that
the data for $A_u^{\rm SFD}>2$, where a significant departure is observed, are
derived from the counts whose faint end extends beyond the incompleteness
limit, in addition to the problem of photometry discussed above. We
take the results with the $u$ band hereinafter only for the purpose to
see the broad, rather than quantitative, consistency.

\begin{figure}
\plotone{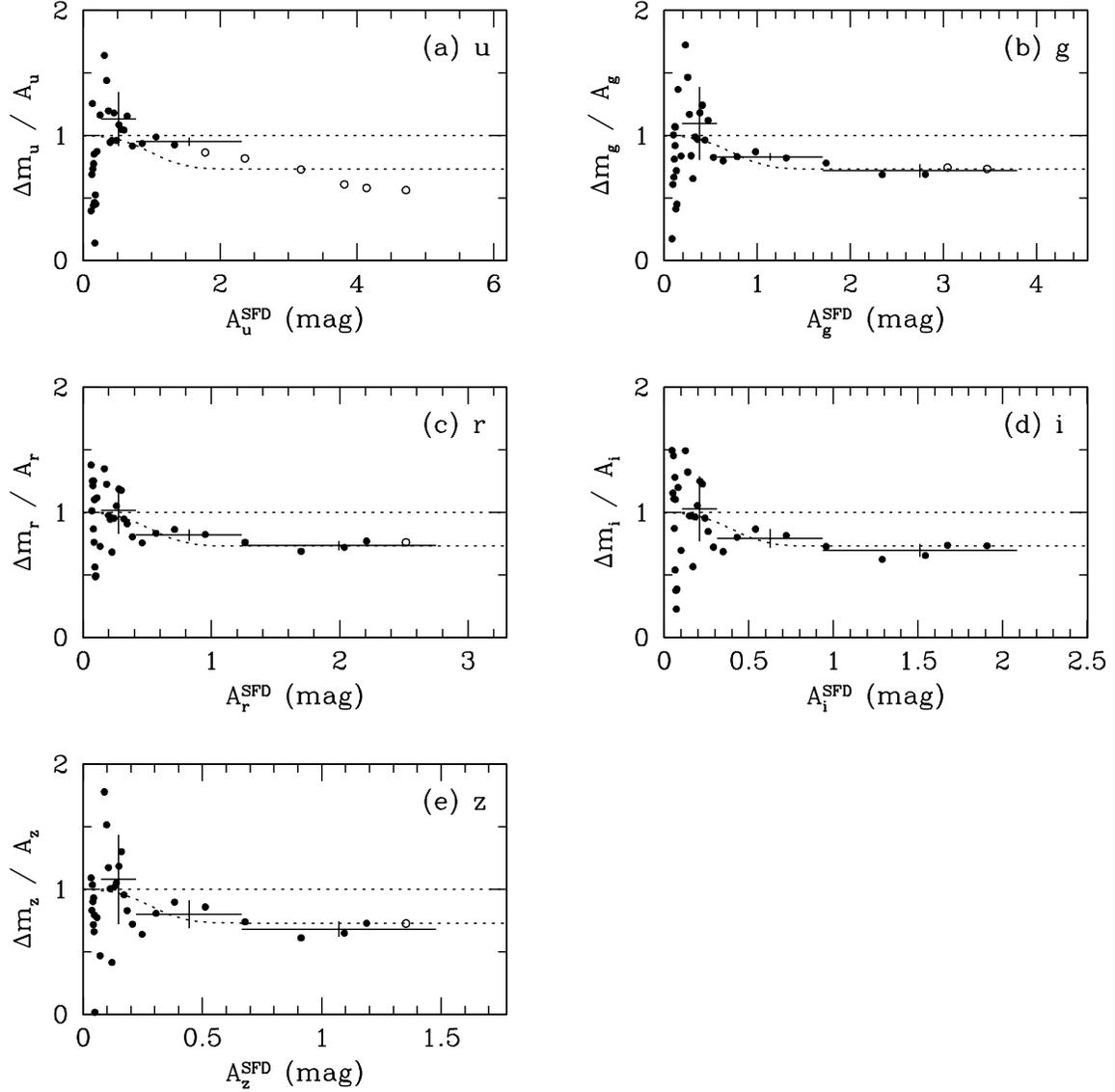}
\caption{The ratio of $\Delta m_\lambda / A_\lambda^{\rm SFD}$ as a
function of $A_\lambda^{\rm SFD}$ for five colour bands. The meaning of open
circles is the same as in Figure \ref{fig:Ar}. The mean values 
for different ranges of $E(B-V)$ ($0.05-0.15$, $0.15-0.45$,
and $0.45-1.00$) are indicated as horizontal bars, and the empirical 
fitting function eq. (\ref{eq:correct}) is shown with dotted curves.
\label{fig:ratio}}
\end{figure}

The departure from the identity regression is more clearly
demonstrated in Figure \ref{fig:ratio}, where the ratio of $\Delta
m_\lambda / A_\lambda^{\rm SFD}$ is plotted as a function of
$A_\lambda^{\rm SFD}$ for each band. The mean values are presented as
horizontal bars (for numerical values see Table \ref{tbl:ratio}) for
the ranges of $E(B-V)=0.05-0.15$, $0.15-0.45$, and $0.40-1.00$. For 
$0.05 < E(B-V) < 0.15$, these ratios are consistent with unity
within $1\sigma$ errors. Apparent departure from unity, 
$\Delta m_\lambda<A_\lambda^{\rm SFD}$, is visible for higher extinction
ranges. The ratio $A_\lambda^{\rm SFD} / \Delta
m_\lambda$ is approximately $1.25$ for $E(B-V)=0.15-0.45$,
and 1.4 for $E(B-V)=0.45-1.00$ irrespective of the colour bands
(again except for the $u$ band). 

\begin{figure}
\plotone{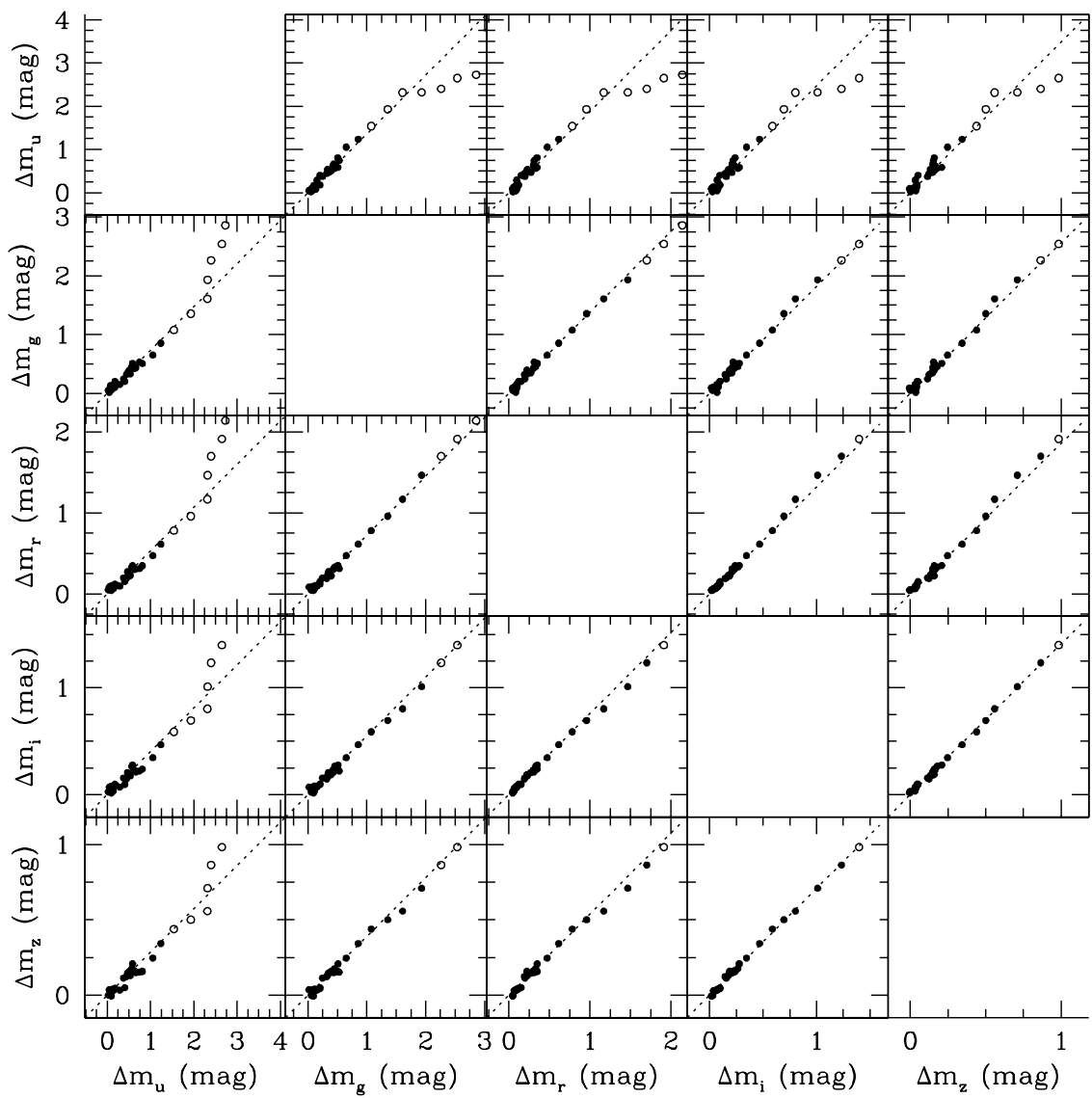}
\caption{The regression of $\Delta m_\lambda$ between different
combinations of $u$, $g$, $r$, $i$, and $z$. The meaning of open
circles is the same as in Figure \ref{fig:Ar}. The dotted lines are
relations expected for the standard extinction law.
\label{fig:ratio2}}
\end{figure}

Two alternative interpretations for this observed departure are (1) the
reddening function $k(\lambda)$ is non-linear for a large $A_\lambda$
or (2) the selective reddening $E(B-V)$ estimated by SFD is not correct
for large extinction. To distinguish between the two possibilities,
the regressions of $\Delta m_\lambda$ among different colour bands are
plotted in Figure \ref{fig:ratio2}. The dotted lines show the
relations expected for the standard extinction law, i.e., $\Delta
m_{\lambda_1}/\Delta m_{\lambda_2}=k(\lambda_1)/k(\lambda_2)$. Note
that the scales of the abscissa and ordinate are not identical. The
figure shows that the data points follow the expected relation; no
non-linearity is observed in the extinction curve across the regions
of low to high extinction.

\begin{figure}
\plotone{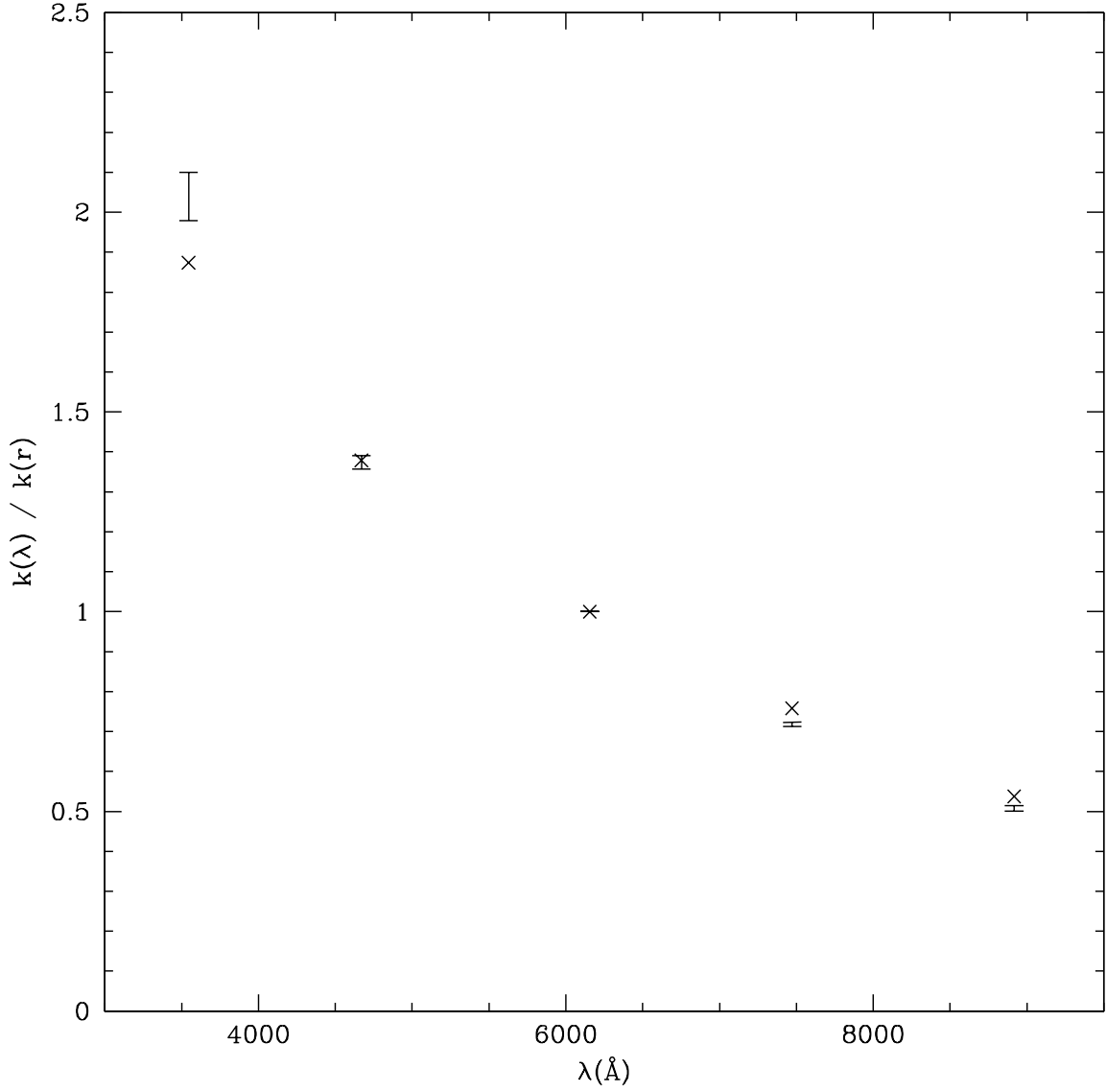}
\caption{$k(\lambda)/k(r)$ obtained from galaxy number counts
as  $\Delta
m_\lambda/\Delta m_r$ (data with error bars) and
those from the standard extinction curve (crosses).
\label{fig:kk}}
\end{figure}

We may estimate $k(\lambda)$ relative to the reference band, which we
take to be the $r$ band. Defining $k'_\lambda = k(\lambda)/k(r)$,
we can calculate this quantity from the observed relation $\Delta
m_\lambda = k'_\lambda \cdot \Delta m_r$. The results from
this analysis are shown in Figure \ref{fig:kk}. The $k$
values in Table \ref{tbl:kk} are obtained by multiplying $k(r)=2.751$
on $k'_\lambda$. The observed values of $k(\lambda)$ are consistent
with those of the standard extinction curve.

This analysis suggests that $E(B-V)$ of SFD is
overestimated for high extinction regions.  Using $\Delta m_\lambda /
A_\lambda^{\rm SFD}$ from $g$ to $z$ bands, we suggest the true
selective reddening being written as
\begin{equation}
E(B-V)^{\rm true}=E(B-V)^{\rm SFD}\left[0.87-0.13~ {\rm Erf}\left(\frac{E(B-V)^{\rm SFD}-0.19}{0.11}\right)\right],
\label{eq:correct}
\end{equation}
in terms of $E(B-V)^{\rm SFD}$. The prediction of this function is
presented in Figure \ref{fig:ratio}. 

In agreement with the analysis of \citet{AGa} we ascribe the
overestimate of $E(B-V)$ by SFD to the inaccuracy of the conversion of
$100\mu m$ emission of dust ($D^T$) to selective reddening for high
extinction regions. \citetalias{SFD} assume a simple linear relation
between reddening and $100\mu m$ emission as $E(B-V) = pD^T$, where
$p$ is a parameter.
To determine $p$, \citetalias{SFD} use the relation between the
intrinsic \bv colour and the Mg$_2$ line index of elliptical galaxies.
Of the 389 elliptical galaxies used,
only $\approx20$ reside in regions with $E(B-V) > 0.15$, and there are
no elliptical galaxies in the region of $E(B-V) > 0.4$. We notice that
the fitting of $100\mu m$ vs $E(B-V)$ relation of SFD (see their Figure
6) starts deviating from linear relation for $E(B-V)\gtrsim
0.15$. From their Figure 6, the value of $\delta (B-V)$ is about
$-0.08$ mag for elliptical galaxies whose $E(B-V)$s are between 0.2 and
0.4; this corresponds to overestimation of reddening by $1.2-1.4$ when
a linear relation is assumed. This is quantitatively consistent with
our result. The relation between the reddening and $100\mu m$ emission
should be modified to incorporate the non-linearity.

We briefly discuss the possible reasons for the departure between
$100\mu m$ emission and reddening from a linear
relation. Overestimation of reddening can be caused by the
overestimation of $100\mu m$ emission. According to the procedure of
\citetalias{SFD}, $100\mu m$ emission will be overestimated when color
temperature estimated from the ratio of intensities at 100 and 240
$\mu m$ is underestimated. If dense clouds have larger dust grains,
their equilibrium temperature will be lower even in the same radiation
field. This will cause overestimation of $100\mu m$ emission. Could
the composition be different? \citetalias{SFD} assumed the emissivity
model of $\epsilon_\nu = \nu^\alpha$ with $\alpha = 2.0$. If there are
materials with $\alpha = 1.5$, their temperature will be
underestimated. This is the same sense as we have seen. How about
mixed temperature along the line of sight? From Figure 2 of
\citetalias{SFD} the column density will be underestimated when there
are two regions at different equilibrium temperatures. This is the
opposite sense to what we have seen. From these brief discussion, the
difference of grain size and emissivity model in dense clouds could be
the cause of overestimation of 100$\mu m$ emission. However, we need
more detailed observations to explore the ways of modifying the
\citetalias{SFD} map.

\section{Summary}

We have tested Galactic extinction using galaxy number counts at the
low Galactic latitude using the SDSS galaxy sample of
\citet{OrionDR} that covers 200 deg$^2$ from low $E(B-V)\approx 0.03$
to high $E(B-V)\approx 1$ extinction regions. The variation of galaxy
number counts is consistent with Galactic extinction described by the
prediction of SFD for low extinction regions of $E(B-V) < 0.15$. For
high extinction regions of $E(B-V) > 0.15$, the SFD extinction
prescription overestimates the reddening by a factor up to $1.4$,
which we interpret as a result of the departure of the linear
relation between 100$\mu m$ infrared emission and 
selective extinction.

\acknowledgments

We thank Jill Knapp for invaluable suggestions for improving the
analysis.

Funding for the SDSS and SDSS-II has been provided by the Alfred
P. Sloan Foundation, the Participating Institutions, the National
Science Foundation, the U.S. Department of Energy, the National
Aeronautics and Space Administration, the Japanese Monbukagakusho, the
Max Planck Society, and the Higher Education Funding Council for
England. The SDSS Web Site is {\tt http://www.sdss.org/}.

The SDSS is managed by the Astrophysical Research Consortium for the
Participating Institutions. The Participating Institutions are the
American Museum of Natural History, Astrophysical Institute Potsdam,
University of Basel, Cambridge University, Case Western Reserve
University, University of Chicago, Drexel University, Fermilab, the
Institute for Advanced Study, the Japan Participation Group, Johns
Hopkins University, the Joint Institute for Nuclear Astrophysics, the
Kavli Institute for Particle Astrophysics and Cosmology, the Korean
Scientist Group, the Chinese Academy of Sciences (LAMOST), Los Alamos
National Laboratory, the Max-Planck-Institute for Astronomy (MPIA),
the Max-Planck-Institute for Astrophysics (MPA), New Mexico State
University, Ohio State University, University of Pittsburgh,
University of Portsmouth, Princeton University, the United States
Naval Observatory, and the University of Washington.


\begin{deluxetable}{ccccc}
\tablecolumns{5}
\tablewidth{0pt}
\tablecaption{Parameters for our analysis \label{tbl:param}}
\tablehead{
\colhead{Band} & \colhead{Range of counts} & \colhead{Mag range} & \colhead{Num. of gal} & \colhead{Limiting mag}\\
\colhead{} & \colhead{log $N(m)$ } & \colhead{} & \colhead{} & \colhead{}
}
\startdata
$u$ & $1.0-1.8$ & $18.0-20.0$ & $800$ & $21.6$ \\
$g$ & $1.3-2.1$ & $17.5-19.5$ & $1600$ & $22.4$ \\
$r$ & $1.8-2.6$ & $17.5-19.5$ & $4100$ & $22.0$ \\
$i$ & $1.8-2.6$ & $17.0-19.0$ & $3800$ & $21.2$ \\
$z$ & $1.8-2.6$ & $16.5-18.5$ & $3100$ & $19.8$ \\
\enddata
\end{deluxetable}

\begin{deluxetable}{cccc}
\tablecolumns{4}
\tablewidth{0pt}
\tablecaption{Mean values of $\Delta
m_\lambda / A_\lambda^{\rm SFD}$ \label{tbl:ratio}}
\tablehead{
\colhead{} & \multicolumn{3}{c}{Range of $E(B-V)$} \\
\cline{2-4} \\
\colhead{Band} & \colhead{$0.05-0.15$} & \colhead{$0.15-0.45$} & \colhead{$0.45-1.00$}
}
\startdata 
$u$ & $1.130 \pm 0.063$ & $0.950 \pm 0.019$ & \nodata \\
$g$ & $1.095 \pm 0.085$ & $0.829 \pm 0.015$ & $0.718 \pm 0.031$ \\
$r$ & $1.017 \pm 0.055$ & $0.818 \pm 0.022$ & $0.734 \pm 0.019$ \\
$i$ & $1.028 \pm 0.075$ & $0.791 \pm 0.038$ & $0.694 \pm 0.023$ \\
$z$ & $1.079 \pm 0.103$ & $0.800 \pm 0.057$ & $0.681 \pm 0.031$ \\
\enddata
\end{deluxetable}

\begin{deluxetable}{ccl}
\tablecolumns{3}
\tablewidth{0pt}
\tablecaption{The value of $k(\lambda)$ \label{tbl:kk}}
\tablehead{
\colhead{Colour band} & \colhead{Standard extinction curve} & \colhead{Derived
from $\Delta N(m)$}
}
\startdata
$u$ & 5.155 & $5.611 \pm 0.165$ \\
$g$ & 3.793 & $3.779 \pm 0.045$ \\
$r$ & 2.751 & 2.751 (normalisation) \\
$i$ & 2.086 & $1.9748 \pm 0.015$ \\
$z$ & 1.479 & $1.397 \pm 0.019$ \\
\enddata
\end{deluxetable}

\newpage

\end{document}